\documentstyle[12pt]{article}

\catcode`\@=11
\long\def\@makefntext#1{
\protect\noindent \hbox to 3.2pt {\hskip-.9pt
$^{{\ninerm\@thefnmark}}$\hfil}#1\hfill}		

\def\@makefnmark{\hbox to 0pt{$^{\@thefnmark}$\hss}}  

\def\ps@myheadings{\let\@mkboth\@gobbletwo
\def\@oddhead{\hbox{}
\rightmark\hfil\ninerm\thepage}
\def\@oddfoot{}\def\@evenhead{\ninerm\thepage\hfil
\leftmark\hbox{}}\def\@evenfoot{}
\def\sectionmark##1{}\def\subsectionmark##1{}}

\renewcommand{\thefootnote}{\fnsymbol{footnote}}

\newcounter{sectionc}\newcounter{subsectionc}\newcounter{subsubsectionc}
\renewcommand{\section}[1] {\vspace*{0.6cm}\addtocounter{sectionc}{1}
\setcounter{subsectionc}{0}\setcounter{subsubsectionc}{0}\noindent
	{\normalsize\bf\thesectionc. #1}\par\vspace*{0.4cm}}
\renewcommand{\subsection}[1] {\vspace*{0.6cm}\addtocounter{subsectionc}{1}
	\setcounter{subsubsectionc}{0}\noindent
	{\normalsize\it\thesectionc.\thesubsectionc. #1}\par\vspace*{0.4cm}}
\renewcommand{\subsubsection}[1]
{\vspace*{0.6cm}\addtocounter{subsubsectionc}{1}
	\noindent {\normalsize\rm\thesectionc.\thesubsectionc.\thesubsubsectionc.
	#1}\par\vspace*{0.4cm}}

\newcounter{appendixc}
\newcounter{subappendixc}[appendixc]
\newcounter{subsubappendixc}[subappendixc]

\renewcommand{\appendix}[1] {\vspace*{0.6cm}
        \refstepcounter{appendixc}
        \setcounter{figure}{0}
        \setcounter{table}{0}
        \setcounter{equation}{0}
        \renewcommand{\thefigure}{\Alph{appendixc}.\arabic{figure}}
        \renewcommand{\thetable}{\Alph{appendixc}.\arabic{table}}
        \renewcommand{\theappendixc}{\Alph{appendixc}}
        \renewcommand{\theequation}{\Alph{appendixc}.\arabic{equation}}
        \noindent{\bf Appendix \theappendixc #1}\par\vspace*{0.4cm}}

\def\abstracts#1{{

\centering{\begin{minipage}{12.2truecm}\footnotesize\baselineskip=12pt\noindent
	\centerline{\footnotesize ABSTRACT}\vspace*{0.3cm}
	\parindent=0pt #1
	\end{minipage}}\par}}

\renewenvironment{thebibliography}[1]
	{\begin{list}{\arabic{enumi}.}
	{\usecounter{enumi}\setlength{\parsep}{0pt}
\setlength{\leftmargin 1.25cm}{\rightmargin 0pt}
	 \setlength{\itemsep}{0pt} \settowidth
	{\labelwidth}{#1.}\sloppy}}{\end{list}}

\newcounter{itemlistc}
\newcounter{romanlistc}
\newcounter{alphlistc}
\newcounter{arabiclistc}

\newcommand{\fcaption}[1]{
        \refstepcounter{figure}
        \setbox\@tempboxa = \hbox{\footnotesize Fig.~\thefigure. #1}
        \ifdim \wd\@tempboxa > 6in
           {\begin{center}
        \parbox{6in}{\footnotesize\baselineskip=12pt Fig.~\thefigure. #1}
            \end{center}}
        \else
             {\begin{center}
             {\footnotesize Fig.~\thefigure. #1}
              \end{center}}
        \fi}

\newcommand{\tcaption}[1]{
        \refstepcounter{table}
        \setbox\@tempboxa = \hbox{\footnotesize Table~\thetable. #1}
        \ifdim \wd\@tempboxa > 6in
           {\begin{center}
        \parbox{6in}{\footnotesize\baselineskip=12pt Table~\thetable. #1}
            \end{center}}
        \else
             {\begin{center}
             {\footnotesize Table~\thetable. #1}
              \end{center}}
        \fi}

\def\@citex[#1]#2{\if@filesw\immediate\write\@auxout
	{\string\citation{#2}}\fi
\def\@citea{}\@cite{\@for\@citeb:=#2\do
	{\@citea\def\@citea{,}\@ifundefined
	{b@\@citeb}{{\bf ?}\@warning
	{Citation `\@citeb' on page \thepage \space undefined}}
	{\csname b@\@citeb\endcsname}}}{#1}}

\newif\if@cghi
\def\cite{\@cghitrue\@ifnextchar [{\@tempswatrue
	\@citex}{\@tempswafalse\@citex[]}}
\def\citelow{\@cghifalse\@ifnextchar [{\@tempswatrue
	\@citex}{\@tempswafalse\@citex[]}}
\def\@cite#1#2{{$\null^{#1}$\if@tempswa\typeout
	{IJCGA warning: optional citation argument
	ignored: `#2'} \fi}}

 1
 1
 1

\font\ninerm=cmr9

\begin{document}

\vspace*{-1.25in}
\begin{flushright}
FERMILAB-Conf-95/373-T\\
hep-ph/9511468\\
November 29, 1995\\
\vspace*{0.25in}
\end{flushright}

\centerline{\normalsize\bf TOP QUARK PAIR PRODUCTION:}
\baselineskip=16pt
\centerline{\normalsize\bf SENSITIVITY TO NEW PHYSICS
\footnote{Invited talk at the International Symposium on Heavy Flavor
and Electroweak Theory, Beijing, China, August 17-19, 1995.}
}
\baselineskip=18pt

\centerline{\normalsize STEPHEN PARKE}
\baselineskip=13pt
\centerline{\footnotesize parke@fnal.gov}
\baselineskip=18pt
\centerline{\footnotesize\it Theoretical Physics Department}
\baselineskip=12pt
\centerline{\footnotesize\it Fermi National Accelerator Laboratory
\footnote{\baselineskip=16pt Fermilab is operated by the Universities
Research Association under 
contract with the United States Department of Energy.}
}
\baselineskip=12pt
\centerline{\footnotesize\it P.O. Box 500, Batavia, IL 60510, U.S.A.}

\vspace*{0.5cm}
\abstracts{The production cross--section and distributions
of the top quark are sensitive to
new physics, e.g., the $t\overline{t}$ system can be
a probe of new resonances or gauge bosons that are strongly
coupled to the top quark, in analogy to Drell--Yan production.
The existence of such new physics is expected in
dynamical electroweak symmetry breaking schemes,
and associated with the large mass of the top quark.
The total top production cross--section can be
more than doubled, and distributions significantly
distorted with a  chosen scale
of new physics of $\sim 1 $ TeV in the vector color
singlet or octet $s$--channel.
New resonance physics is most readily discernible in the high--$p_T$
distributions of the single top quark, of the $W$ boson and the mass
distribution of the $t\bar{t}$ pair.
}

\normalsize\baselineskip=15pt
\setcounter{footnote}{0}
\renewcommand{\thefootnote}{\alph{footnote}}

\section{Summary}
Top quark production at the Fermilab Tevatron probes very high mass scales,
$ {\cal{O}}(500 ~GeV)$, and therefore is sensitive to
new physics at this scale.
Hence, it is important that we studied this process with high precision
and compare the results with the Standard Model predictions.

The dominant mode of top quark production at hadron colliders is via
quark-antiquark annihilation or gluon-gluon fusion,
\begin{eqnarray*}
q~\bar{q} & \rightarrow & t ~\bar{t} \\
g~g  & \rightarrow & t ~\bar{t}.
\end{eqnarray*}
Fig. 1 is the QCD cross section for $m_{t} ~=~ 175~GeV$ verses $\sqrt{s}$
for both proton-proton colliders and for proton-antiproton colliders.
At $\sqrt{s} ~=~ 1.8~TeV$ the gluon-gluon fusion is only 10\% of the cross
section for a proton-antiproton collider, the Tevatron,
whereas at a $14~TeV$ proton-proton
collider, the LHC, gluon-gluon fusion is 90\% of the cross section.

\begin{figure}[hbt]
\vspace{8cm}
\includegraphics{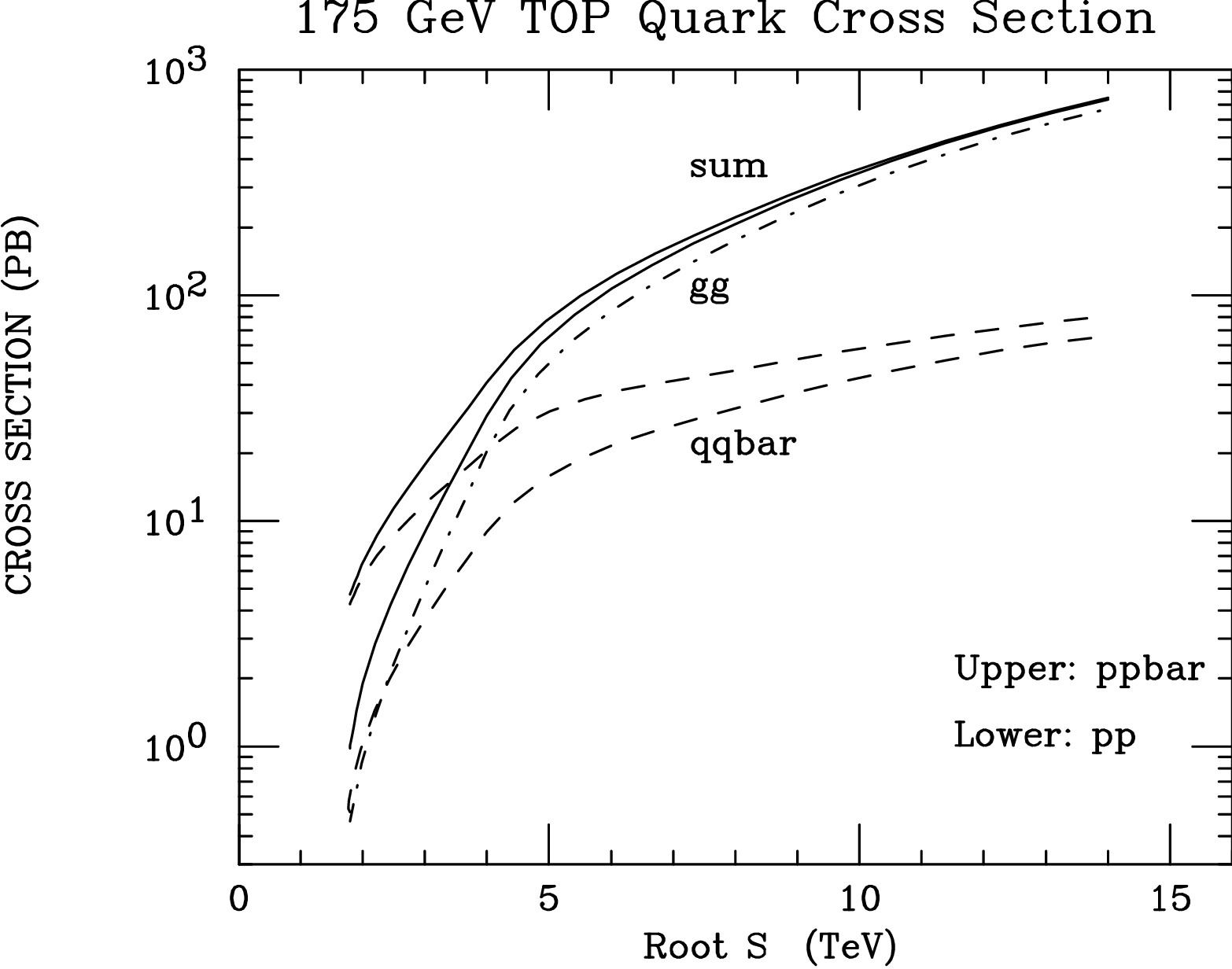}
\vspace{-1.5cm}
\caption[]{
QCD Top quark Production cross section as a function of $\sqrt{s}$,
for quark-antiquark annihilation (dashes), gluon-gluon fusion (dot-dash)
and the sum (solid)  for both proton-antiproton (upper)
and proton-proton (lower) colliders.}
\label{xsec}
\end{figure}

\begin{figure}[hb]
\vspace{7cm}
\includegraphics{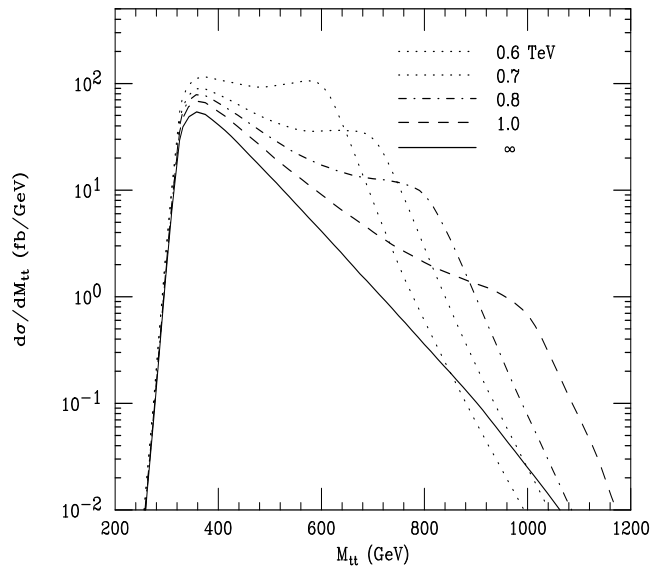}
\vspace{-1.5cm}
\caption[]{The invariant mass of the $t\bar{t}$ pair for
the topcolor octet model.}
\label{newdyna}
\end{figure}

Hill and Parke~\cite{hp} have studied the effects of new physics
on top quark production in a general operator formalism as well as in
topcolor models. In these models the distortions in top quark production
and shape are due to new physics in the $q\bar{q}$ subprocess.
The effects of a coloron which couples weakly to the light generations
but strongly to the heavy generation is given in Fig. 2.

\begin{figure}[hbt]
\vspace{9cm}
\includegraphics{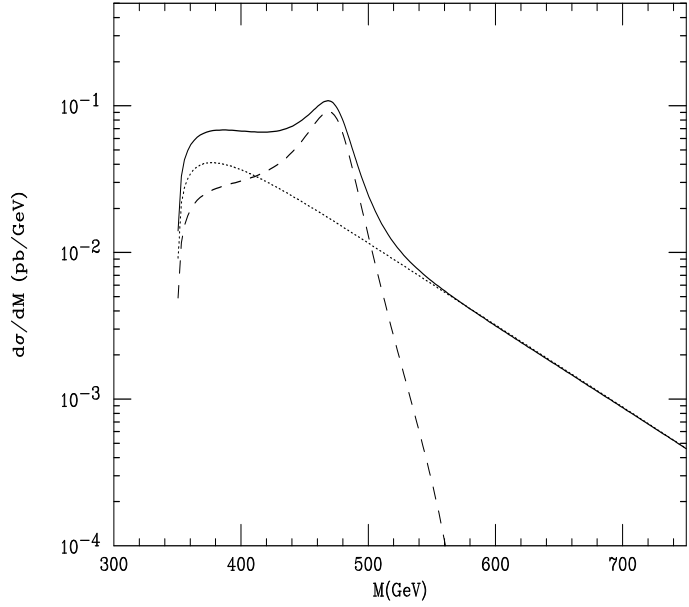}
\vspace{-1.5cm}
\caption[]{The invariant mass of the $t\bar{t}$ pair for
the two scale technicolor model.}
\label{newdynb}
\end{figure}

\newpage
Similarly Eichten and Lane~\cite{el} have studied the effects of
multi-scale technicolor on top production through
the production of a techni-eta
resonance, see Fig. 3.
Here the coupling of the techni-eta is to $gg$, therefore only
this subprocess is different than the standard model.
At the Fermilab Tevatron top production is dominated by $q\bar{q}$ annihilation
while at the LHC it is the $gg$ fusion subprocess that dominates. Therefore
these models predict very different consequences for top production at the
LHC.
\newpage

Therefore the top quark is an exciting new window on
very high mass scale physics.
While exploring the vista from this window we should be on the lookout for
any deviation from the Standard Model which will provide us with information
about that elusive beast, the mechanism of electro-weak symmetry breaking.
Because the mass of the top quark is very heavy,
this quark is the particle most strongly
coupled to the electro-weak symmetry breaking sector.
Therefore the deviations could be seen at zeroth
order or may require more subtle measurements.
What is needed is hundreds of top-antitop pairs as soon as possible.
Then, watch out for surprises!

\section{Acknowledgements}

I would like to thank E.~Eichten, K.~Lane and especially
my collaborator C.~Hill for many discussions on this topic.

\end{document}